\documentstyle[prb,aps,epsfig]{revtex}
\begin{document}
\draft

\twocolumn[\hsize\textwidth\columnwidth\hsize\csname@twocolumnfalse\endcsname

\title{Dynamics of a three terminal mechanically flexible tunneling contact}

\author{A. Isacsson}
\address{Department of Applied Physics, Chalmers University of Technology 
and G\"oteborg University}
\address{S-412 96 G\"oteborg, Sweden}

\date{\today}
\maketitle
\begin{abstract}
The dynamics of a nanoelectromechanical system in the form of a 
three-terminal tunneling device
is studied by analytical and numerical methods. 
The main results are the existence of bistable stationary states resulting in
directly detectable chaotic behavior. 
\end{abstract}

%
]
\narrowtext

\section{Introduction}
The growing interest in nano electromechanical
systems (NEMS) is partly due to the advances in 
microfabrication by self-assembly of biometallic composites
where metallic or semiconducting clusters are combined with organic molecules
such as polymers or DNA \cite{Andres,Sivan,DNAsensor,DNA1,DNA2}. 
Since the organic chain molecules used in self-assembly
are typically a few orders of magnitude softer \cite{alvarez,smith1,smith2}
than ordinary solids their utilization in nanoelectronics implies that 
mechanical degrees of freedom
can have great impact on the electronic transport properties in 
such systems. 

We have previously\cite{Gorelik,Isacsson} investigated how the mechanical 
degrees of freedom couple to the electrical transport properties
in a Coulomb blockade double junction where the central island is free to
move in a parabolic potential. A dynamical instability was shown to exist 
in this structure that caused the central island to oscillate between the 
external electrodes. In the low temperature limit, where charging
effects are important, this instability gave rise to
a mechanically mediated current that approached the value $I=2eNf$
where $e$ is the elementary charge, $N$ the maximum number of excess electrons
allowed on the island and $f$ the frequency of elastic vibrations. 
The fluctuations of 
this current have been shown to disappear exponentially with
decreasing temperatures \cite{Weiss} implying that this system may be well 
suited for 
current standard purposes. In realizing a current standard device of this type
a grain placed on a flexible cantilever
positioned between the two electrodes can replace 
the grain and the soft molecular links considered in\cite{Gorelik,Isacsson}.
Such nanoscale mechanical resonators have 
successfully been fabricated  
\cite{blick1,blick2}. It has also been
suggested by Tuominen {\it et al.}\cite{krotkov}, who studied a macroscopic
electromechanical system at room temperature, that carbon 
nanotubes connected to nanoscale metallic grains\cite{liu,poncharal}, 
could be used for this purpose.

\begin{figure}
\epsfig{file=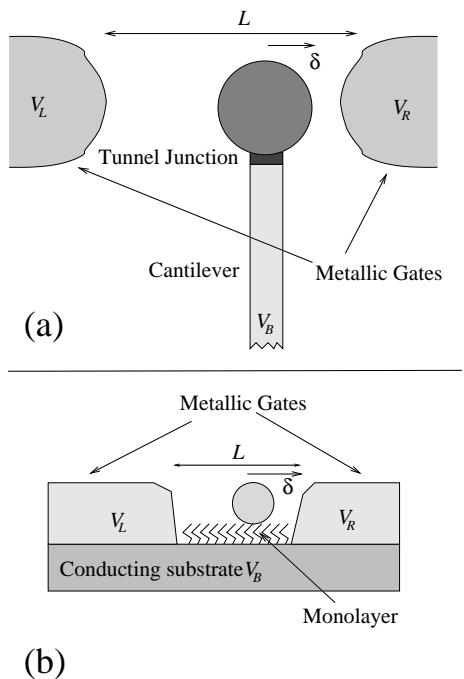, width=7cm}
\caption{Schematic layout of the system. (a) A cantilever attached to a 
conducting grain through a tunnel junction situated asymmetrically between two 
metallic electrodes separated by a distance $L$. The electrodes and the 
cantilever are biased with voltages $V_L$, $V_R$ and $V_B$ respectively.
The distance between the grain and the electrodes are such that electrons
may tunnel through the structure.
(b) Self assembled single electron transistor structure. The dynamical
features discussed for system (a) are relevant also for this system.}
\label{fig:cantileverthing}
\end{figure}

In this paper it is shown that when a system consisting of
a grain attached to the tip of a cantilever by a tunnel
junction (see Fig.~\ref{fig:cantileverthing}a)
is treated as a three terminal device it shows
a rich dynamical structure ranging from stable limit cycle
behavior to deterministic chaos. Since this investigation focuses
on the dynamical properties, the temperature is considered to
be large enough for charging effects to be absent while
still low enough for thermal fluctuations to be negligible.
\noindent
The same model also applies to the dynamics of the single electron transistor
structure in Fig.~\ref{fig:cantileverthing}b. The central 
island is here resting on a mechanically soft insulating monolayer 
allowing for center of mass motion of the grain. This system should
of course be studied in the low temperature limit where charging effects 
are of importance. However, since much of the dynamical features of the system in
the high temperature limit are expected 
to survive at lower temperatures the results obtained in this work is 
of interest for further studies in the low temperature regime.    
\section{Model system} 
The system, as depicted in Fig.~\ref{fig:cantileverthing}a 
consists of a small metallic grain, typically a few nanometers in diameter, 
attached to the tip of a metallic cantilever 
through a tunnel junction. 
The other end of the cantilever is assumed to be
clamped and connected to a voltage source $V_B$. The grain attached to the 
cantilever 
is situated between two
metallic electrodes located close enough to the grain to allow 
tunneling but with tunneling resistances
much larger than that of the grain-cantilever junction.  
If the typical frequency of the dynamical motion of the system is well
below the plasmafrequency of the metallic components the charges
and potentials on the conductors are related by the matrix $C_{ij}$;
\begin{figure}
\epsfig{file=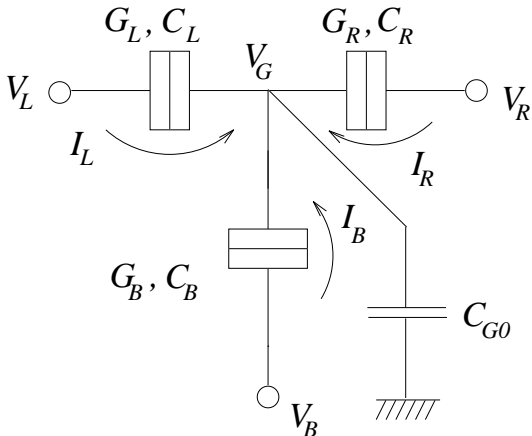, width=7cm}
\caption{Electric circuit model. Each tunnel junction is modeled by a capacitor
in parallel with a tunnel resistance $R_{L,R,B}=G^{-1}_{L,R,B}$. Furthermore
the grain-ground capacitance has been taken into account through the capacitance
$C_{G0}$. Only the ``nearest neighbor'' capacitances are considered in this model.}
\label{fig:x01}
\end{figure}\noindent
$$Q_i=\sum\limits_jC_{ij}V_j.$$
The system can be mapped onto an electrical network model seen in 
Fig.~\ref{fig:x01}. In this model only ``nearest neighbor''
coefficients in $C_{ij}$ have been taken into account.
Each tunnel junction is modeled by a resistance $R$ in parallel 
with a capacitance $C$ the latter being related to the coefficeints $C_{ij}$ 
by linear transformations.
Using this model, the grain potential $V_G$ can be expressed in terms of the 
grain charge $Q_G$ and the applied bias voltages $V_{R,L,B}$ 
(cnf. Fig.~\ref{fig:cantileverthing}) as
$$V_G(Q_G)=\frac{Q_G+V_RC_R+V_LC_L+V_BC_B}{C_{\Sigma}},$$
where $C_{\Sigma}=C_L+C_R+C_B+C_{G0}$.
The currents $I_{L,R,B}$ flowing from the electrodes
to the grain, as shown in Fig.~\ref{fig:x01}, determine an equation of 
motion for $Q_G$;
\begin{displaymath}
\frac{{\rm d}Q_G}{{\rm d}t}=I_L+I_R+I_B.
\end{displaymath}
Using $I_{L,R,B}=(V_{L,R,B}-V_G)G_{L,R,B}$
where $G_{L,R,B}$ are conductances of the tunnel junctions 
(cnf. Fig.~\ref{fig:x01}) one finds
$$\frac{{\rm d}Q_G}{{\rm d}t}=V_LG_L+V_RG_R+V_BG_B-V_G(G_L+G_R+G_B).$$

The flexibility of the cantilever allows for a mechanical degree of freedom in 
the system. 
We will assume that 
this freedom is one-dimensional, i.e. that the cantilever only bends in one 
direction and
that this bending is so 
small that we can consider the grain to move on a straight line between the 
two electrodes. The deflection $\delta$ 
of the grain from the equilibrium position towards the right electrode is 
described by the equation of motion
$$\ddot{\delta}+\gamma\dot{\delta}+\omega_0^2\delta={\cal E}Q_G/m,$$ 
where $\omega_0$ is the elastic frequency of the system and $\gamma$
a parameter describing the damping. The mass $m$ appearing on the right hand side
is an effective mass depending on the precise geometry and design of the 
cantilever-grain
system. In the force term only the effect of the 
electro-static field between the left and right electrodes 
${\cal E}\approx(V_L-V_R)/L$ to linear order in $Q_G$ has been considered. 
The tunneling conductances $G_{L,R}$ have a sensitive dependence on the
grain displacement $\delta$
$$G_{L,R}=G_{L,R}^0\exp\left(\mp\frac{\delta}{\lambda}\right),$$ 
where the tunneling length $\lambda$ is determined by the work function 
$\phi$ of the 
electrodes, 
\begin{displaymath}
\lambda\approx\left(\frac{2\sqrt{2m_e\phi}}{\hbar}\right)^{-1}.
\end{displaymath}
The 
variations of the capacitances with position have been neglected since they are  
much smaller than
the changes in the conductances which dominate the nonlinear behavior of the 
system.
By measuring the deflection $\delta$ in units of lambda i.e. 
$\delta=\xi\lambda$, 
one arrives at the following system of equations,
\begin{eqnarray}
\dot{\xi}&=&\Pi \label{dyn1}\\
\dot{\Pi}&=&-\gamma\Pi-\omega_0^2\xi+\left(\frac{V_L-V_R}{Lm\lambda}\right)Q_G 
\label{dyn2}\\
\dot{Q}_G&=&V_LG_L^0e^{-\xi}+V_RG_R^0e^{\xi}+V_BG_B-V_G(Q_G)G_\Sigma \label{QG}
\end{eqnarray}
where
\begin{eqnarray}
G_{\Sigma}(\xi)&=&G_L^0e^{-\xi}+G_R^0e^{\xi}+G_B
\end{eqnarray}
and
\begin{eqnarray}
V_G(Q_G)&=&\frac{Q_G+V_LC_L+V_RC_R+V_B C_{B}}{C_{\Sigma}} \label{VT}\mbox{ }.
\end{eqnarray}

\begin{figure}
\epsfig{file=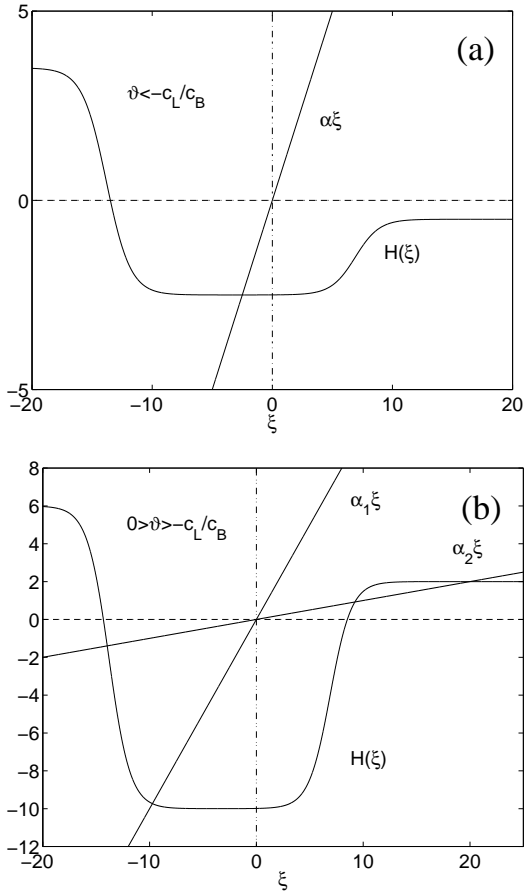, width=7cm}
\caption{Solutions to the fixed point equation $\alpha\xi=H(\xi)$. 
When $\vartheta=V_L/V_B<0$ weither one or three solutions exist depending 
on whether $\vartheta$ is larger or smaller than the ratio $-C_L/C_B$ 
and what value $\alpha\propto\omega_0^2$ assumes. In (a) only one 
stationary point exists while in (b) either one or three stationary points
appear.}
    \label{fig:th1}
\end{figure}\noindent
\section{Fixed Points and stability}
In order to characterize the dynamical behavior of the system the 
existence of fixed points is investigated and then the
stability of these points is considered.
Throughout the rest of the article the right electrode potential will be 
used as reference i.e. we will put $V_R=0$ from here on. With this 
convention the stationary 
points satisfy the system of equations
\begin{equation}
\left\{\begin{array}{lcl}
0&=&\Pi \nonumber \\
0&=&-\gamma\Pi-\omega_0^2\xi+\left(\frac{V_L}{Lm\lambda}\right)Q_G \nonumber \\
0&=&V_LG_L^0e^{-\xi}+V_BG_B-\frac{Q_G+V_LC_L+V_B C_B}{C_{\Sigma}}
G_\Sigma(\xi). 
\end{array}\right.\label{equis}
\end{equation}
Defining $\alpha\equiv\omega_0^2Lm\lambda/C_\Sigma$, 
$\eta^2=V_L^2$ and $\sigma=V_LV_B$, equation (\ref{equis}) can be 
recast in the form $\alpha\xi=H(\xi)$ where
$$H(\xi)=\frac{\eta^2g_Le^{-\xi}+\sigma}{g_Le^{-\xi}+g_Re^{\xi}+1}-
(\eta^2c_L+\sigma c_B).$$
\begin{figure}
\epsfig{file=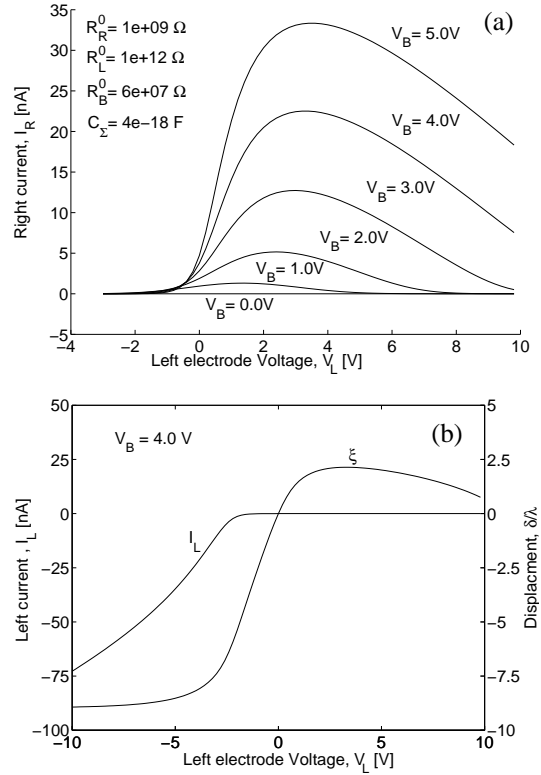, width=7cm}
\caption{Stationary operation. When the 
damping constant $\gamma$ is large 
the static deflection of the cantilever will cause a pronounced transistor-like 
action
due to the exponential decrease of the tunneling resistance between the grain 
and the right electrode. Figure (a) shows the current flowing from the grain 
to the right electrode
as a function of the voltage $V_L$ applied to the left electrode. 
The different curves
correspond to different biases $V_B$ applied to the bottom electrode. 
In (b) the displacement $\delta$ of the cantilever and the current flowing 
from the left electrode to the
grain is plotted for the case $V_B=4.0$ V. }
    \label{fig:st1}
\end{figure}\noindent
Here the dimensionless conductances $g_{R,L}=G_{R,L}^0/G_B$ and
capacitances $c_{L,B}=C_{L,B}/C_\Sigma$ have been introduced.
Since $H(\xi)$ is a restricted function
the equation $\alpha\xi=H(\xi)$ has at least one solution
and at the most three solutions. Defining $\vartheta=\eta^2/\sigma=V_L/V_B$
three different cases can be identified,
\begin{itemize}
\item $\vartheta<0$, $H(\xi)$ has one minimum located at 
$\xi_0=\frac{1}{2}\ln([1-\vartheta]g_L/g_R)$
which means that the system has either one or three fixed points. 
\item $0<\vartheta<1$,  $H(\xi)$ has one maximum located at 
$\xi_0=\frac{1}{2}\ln([1-\vartheta]g_L/g_R)$ and again we may have either one or 
three fixed points.
\item $\vartheta>1$,  In this case $H(\xi)$ 
is monotonic hence only one fixed point can exist.
\end{itemize}
When $\vartheta<0$  two different cases can be distinguished; 
$\vartheta<-c_B/c_L$ and $0>\vartheta>-c_B/c_L$. In the first case 
$\lim_{\xi\rightarrow +\infty}H(\xi)<0$ and 
only one solution lying in the left half plane is possible. In the second 
case the corresponding limit is 
positive and if $\alpha$ is chosen small enough three solution will appear, 
one in the left half plane and 
two in the right. The different scenarios are shown in Fig.~\ref{fig:th1}.
When $0<\vartheta<1$ one can again single out two cases; 
$c_B/(1-c_L)<\vartheta<1$ and $0<\vartheta<c_B/(1-c_L)$. 
When $c_B/(1-c_L)<\vartheta<1$ then  
$\lim_{\xi\rightarrow -\infty}H(\xi)>0$ and only
one solution located in the right half plane is possible. 
In the other case the limit is negative and one can find
two more solutions in the left half plane by choosing $\alpha$ 
sufficiently small.

A stability analysis of the fixed points obtained above shows that:
In the case of only one fixed point this point will be a stable 
node, i.e. all eigenvalues Jacobian of the system are real and negative\cite{Hilborn}, if the 
damping $\gamma$ exceeds the critical damping $\gamma_c$, 
\begin{eqnarray}
\gamma_c&=&-\frac{1}{2}\left(\tau^{-1}(\xi)+\omega_0^2
\tau(\xi)\right.\nonumber \\
&-&\left.\sqrt{(\tau^{-1}(\xi)+\omega_0^2\tau(\xi))^2-
4\frac{\omega_0^2H^\prime(\xi)}{\alpha}}\right).
\end{eqnarray}
Here the total $RC$-time $\tau(\xi)$ is defined as 
\begin{displaymath}
\tau^{-1}(\xi)\equiv \frac{G_\Sigma(\xi)}{C_\Sigma}=
-\frac{\partial \dot{Q}_G}{\partial Q_G}.
\end{displaymath}
In the case of three fixed points one will be conditionally stable depending
on whether $\gamma$ is larger or smaller than $\gamma_c$. 
The second ``middle''  one (cnf. Fig.~\ref{fig:th1}b) will 
not be a node but instead be a saddle point 
of index 1 (two negative and one positive eigenvalue of the Jacobian)
and hence not a stationary point irrespectively of the value 
of $\gamma$ 
while the remaining point (corresponding to the rightmost solution in figure 
\ref{fig:th1}b) will always be a stable node.
\begin{figure}
\epsfig{file=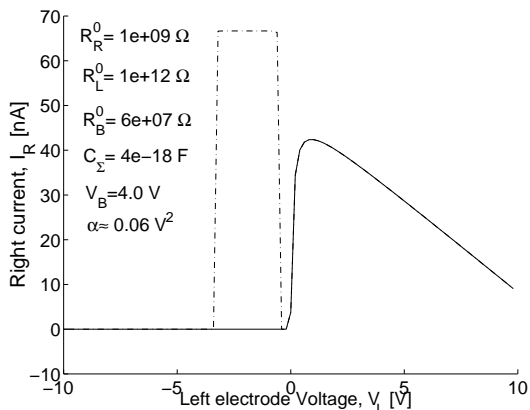, width=7cm}
\caption{Bistable operation. When $\alpha$ is small enough
two stable fixed points emerge leading to a bistable situation. 
The solid line corresponds to the
``expected'' solution while the dashed lines correspond to the new second stable
root 
to the fixed-point equations.}
    \label{fig:st2}
\end{figure}\noindent
\section{Stationary operation}
When $\sqrt{\alpha}$, which is proportional to the square of the frequency of elastic 
vibrations $\omega_0$, 
is large compared to the applied bias voltages one expects to find 
only one stationary solution as discussed above. 
In Fig.~\ref{fig:st1}a
the $I-V$ characteristics for this case is shown for an asymmetric setup with 
$R_R^0=1$ G$\Omega$, $R_L^0=1$ T$\Omega$,
$R_B=6$ M$\Omega$ and all capacitances set to 1 aF. The figure shows the current 
flowing from the grain 
to the right electrode (as opposed to $I_R$ defined in Fig.~\ref{fig:x01} which
was defined in the opposite direction) as a function of the voltage 
applied to the left electrode. The different curves
correspond to different biases applied to the bottom electrode. 
For negative $V_L$ almost no current 
flows in the system since the
grain is essentially disconnected from both leads. As the field is increased 
the \mbox{cantilever} will start to deflect towards 
the right electrode causing an exponential decrease in $R_L$ allowing 
the current to grow.

As $V_L$ is further
increased the charge on the grain will eventually become negative due to the 
capacitive coupling to the left electrode
resulting in a decreased deflection disconnecting it from the leads once again. 
In Fig.~\ref{fig:st1} the the displacement of the grain 
along with the current flowing to it from the left electrode is shown 
as a function of $V_L$ when 
$V_B=4.0$ V. This current is 
essentially zero until a bias of approximately $-3.0$ V is reached. Furthermore, 
from this graph it can be seen that the displacement of
the grain is just a few times $\lambda$ (for Au $\lambda$ is typically 0.5 \AA).

For $\sqrt{\alpha}$ small compared to the applied bias 
voltages, it is possible to have two stable 
fixed points in the system.
Using exactly the same parameters as above but with a reduced $\alpha$ 
this
second solution appears.
In Fig.~\ref{fig:st2} the $I-V$ characteristics of this {\it bistable mode} 
is shown. The solid line
correspond to the same solution as above and the dashed solution to the 
new one that appear due to the reduced $\alpha$.  
\begin{figure}
\epsfig{file=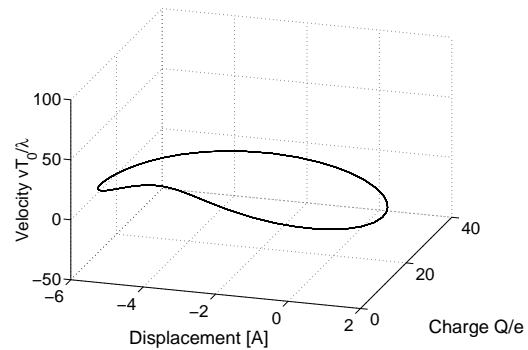, width=7cm}
\caption{Phase space trajectory for a stable limit cycle. As the damping
$\gamma$ is reduced below the critical damping $\gamma_c$ the system settles
in to a stable limit cycle.}
    \label{fig:P1}
\end{figure}\noindent

\section{Dynamical operaton}
In the dynamical regime the system displays a 
rich structure. 
One of the most interesting features is that this is a 
nanoscale system with directly
detectable chaotic behavior. This means that in order to determine what type of 
motion the system exhibits it is sufficient to monitor the currents. 
Due to the multitude of parameters and the system's 
complex dependence on these, only a few archetypical cases will be 
illustrated by means of numerical
integration of the equations of motion (\ref{dyn1})-(\ref{VT}). In the presented
simulations a system with the same parameters as in the 
static case (see the previous section) is considered 
but with a damping rate $\gamma$ reduced below $\gamma_c$. 
The different simulations then correspond to different sets of 
bias voltages $V_B$ and $V_L$.
\begin{figure}
\epsfig{file=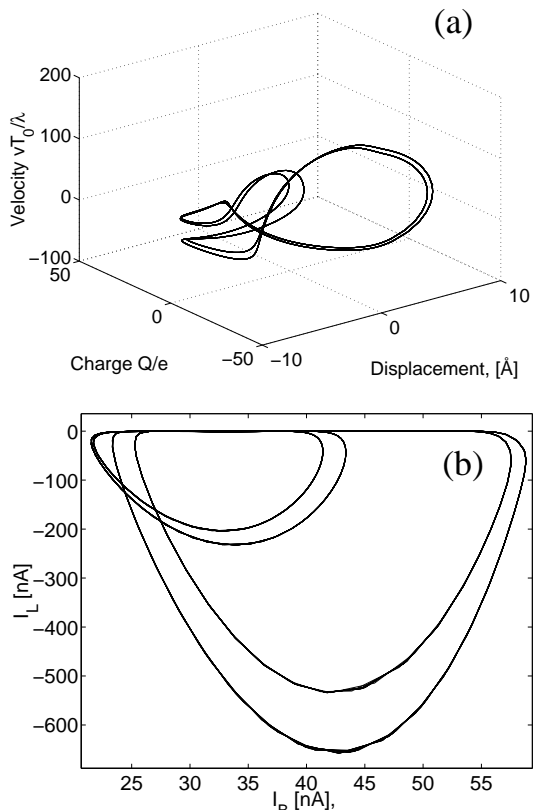, width=7cm}
\caption{Multiply period doubled phase-space trajectory and 
the Corresponding plot of the currents $I_L$ and $I_B$.
(a) Biasing the system towards the situation with three fixed points
will cause subsequent period doublings of the limit cycle. 
(b) By monitoring the currents flowing in the left lead and the bottom lead at
the same time these period doublings can be detected.}
    \label{fig:PXX}
\end{figure}\noindent
\subsection{One Fixed Point}
\begin{figure}
\epsfig{file=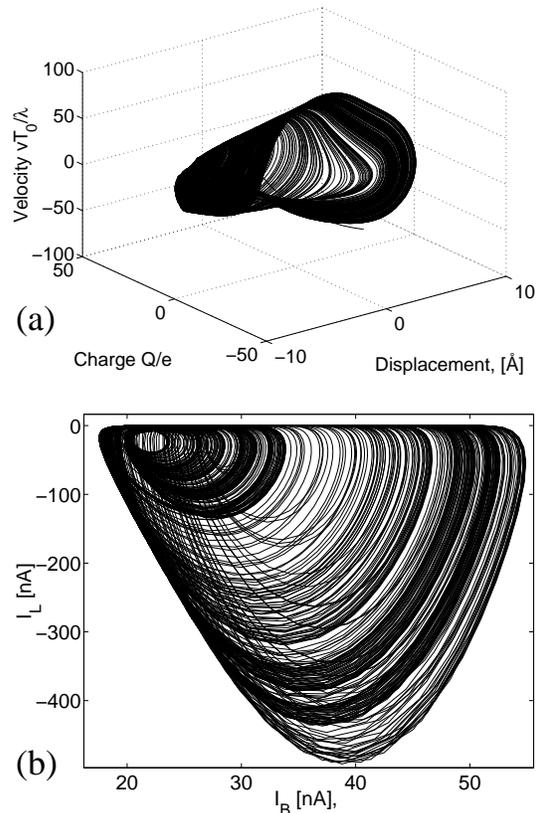, width=7cm}
\caption{Chaotic motion. (a) Phase space trajectory in the chaotic regime.
Biasing the system very close to the situation with three fixed points 
the system becomes chaotic. 
(b) Chaos is also reflected in the corresponding plot of the currents $I_L$ 
and $I_B$.}
    \label{fig:chaos}
\end{figure}

We first consider the situation when the system has only one 
fixed point corresponding
to the situation in Fig.~\ref{fig:th1}a. 
This is achieved by using a fixed
voltage $V_B=3$ V and imposing a positive bias voltage $V_L$.
For small values of this voltage the system remains stable, 
as expected, until a critical
threshold voltage is reached. Further biasing leads, for positive $V_L$, to a 
limit cycle regime. For the range of positive voltages where the algorithm was 
stable
this cycle remained. For negative bias voltages there exists a 
threshold voltage as
well, and as this is reached, a stable limit cycle appears. 
An example of this
type of motion is shown in Fig.~\ref{fig:P1} (recorded at $V_L=-0.35$ V). 
Decreasing $V_L$ moves the system toward the situation in Fig.~\ref{fig:th1}b
i.e. we approach the situation with three fixed points. 
This leads to a sequence of period doublings. The behavior in this regime
is illustrated in Fig.~\ref{fig:PXX}a. These period doublings can be directly
detected by simultaneously measuring the currents $I_B$ and $I_L$ and
plotting them as in Fig.~\ref{fig:PXX}b. Lowering the bias more
eventually leads to a totally chaotic regime like the one in 
Fig.~\ref{fig:chaos} ($V_L=-0.6$ V) which is again reflected in the currents.
\noindent
Further lowering of $V_L$ after this point leads to an alternating
series of period doubled limit cycles and chaotic trajectories until 
three fixed points appear in the system.

\subsection{Three Fixed Points}
\begin{figure}
\epsfig{file=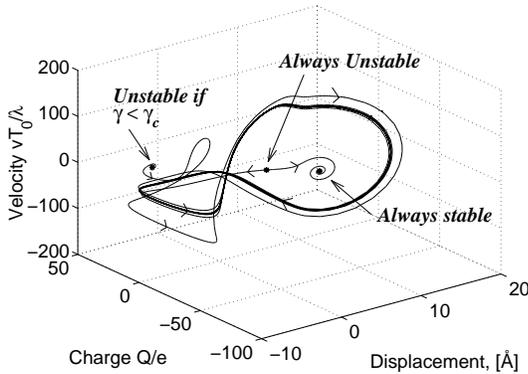, width=7cm}
\caption{Phase space trajectory when three fixed points are present.
When the system has three fixed points, one of them will always be stable,
one will always be unstable and one will be is unstable if $\gamma<\gamma_c$. 
The three fixed points are indicated by * in the figure. The stable limit
cycle that exists in this case can also be seen in the figure.}
    \label{fig:3ps}
\end{figure}

In order to  be able to study the case with three fixed points, $V_B$ 
was raised to 6 V while $V_L$ was set to 0.7 V. This  corresponds to the 
situation with three solutions in Fig.~\ref{fig:th1}b. Numerical integration 
revealed the structure displayed in Fig.~\ref{fig:3ps}. Starting close to the 
conditionally stable fixed point, the leftmost one in Fig.~\ref{fig:th1}b, 
a stable limit cycle is eventually reached.
Starting the simulation in the vicinity of the middle one (always unstable) 
the trajectory
either connects to this limit cycle or becomes attracted by the third fixed 
point, which is always stable. 
\noindent
\section{Conclusions}
We have shown that the three terminal flexible tunneling structures in
Fig.~\ref{fig:cantileverthing}, which are of interest for both current standard
purposes as for self assembled quantum devices, have several characteristic
dynamical features: When the damping $\gamma$ in the system is 
high (low quality factor) the system displays a stationary behavior 
which for some parameter values can be bistable. If the quality factor is 
large enough, i.e. the damping satisfies $\gamma<\gamma_c$, 
the dynamics of the system range from stable limit cycle behavior 
to deterministic chaos. It is furthermore possible to have a situation 
where stable fixed points coexist with stable limit cycles. 
The chaotic motion of the system can be directly detected by
measuring the currents flowing from the terminals.

\acknowledgments
The author would like to acknowledge Leonid Gorelik, Robert Shekhter 
and Sara Blom for fruitful discussions.
This work has been supported by the Swedish Research Council for 
Engineering Sciences (TFR) and the Swedish Strategic Research 
Foundation (SSF) program ``Quantum Devices and Nano-Science''.

\end{document}